\documentclass[aps,prb,twocolumn,groupedaddress]{revtex4}
\usepackage{graphicx,amsmath}
\begin{document}

\title[Grid-dose spreading for dose distribution calculation]{The grid-dose-spreading algorithm for dose distribution calculation in heavy charged particle radiotherapy}
 
\author{Nobuyuki Kanematsu}\email{nkanemat@nirs.go.jp}
\author{Shunsuke Yonai}
\affiliation{Department of Accelerator and Medical Physics, Research Center for Charged Particle Therapy, National Institute of Radiological Sciences, 4-9-1 Anagawa, Inage-ku, Chiba 263-8555, Japan}

\author{Azusa Ishizaki}
\affiliation{Department of Quantum Science and Energy Engineering, Tohoku University, Aramaki-Aza-Aoba 01, Aoba-ku, Sendai 980-8579, Japan}

\date{\today}

\begin{abstract}
A new variant of the pencil-beam (PB) algorithm for dose distribution calculation for radiotherapy with protons and heavier ions, the grid-dose spreading (GDS) algorithm, is proposed.
The GDS algorithm is intrinsically faster than conventional PB algorithms due to approximations in convolution integral, where physical calculations are decoupled from simple grid-to-grid energy transfer.
It was effortlessly implemented to a carbon-ion radiotherapy treatment planning system to enable realistic beam blurring in the field, which was absent with the broad-beam (BB) algorithm.
For a typical prostate treatment, the slowing factor of the GDS algorithm relative to the BB algorithm was 1.4, which is a great improvement over the conventional PB algorithms with a typical slowing factor of several tens.
The GDS algorithm is mathematically equivalent to the PB algorithm for horizontal and vertical coplanar beams commonly used in carbon-ion radiotherapy while dose deformation within the size of the pristine spread occurs for angled beams, which was within 3 mm for a single proton pencil beam of $30^\circ$ incidence, and needs to be assessed against the clinical requirements and tolerances in practical situations.
\end{abstract}



\maketitle

\section{Introduction}

Ceaseless efforts for accuracy improvement and constant progress in computing technology have made the pencil-beam (PB) algorithm be the standard method for dose distribution calculation in heavy charged particle radiotherapy with protons and heavier ions.\cite{Petti 1992, Hong 1996, Deasy 1998, Kanematsu 1998, Schaffner 1999, Russell 2000, Kramer 2000, Szymanowski 2001, Hollmark 2004, Soukup 2005, Kanematsu 2006}
In the PB algorithm, a treatment beam is divided into elementary pencil beams with developing transverse spread as they penetrate through heterogeneous medium to handle spatial modulation of beam scatter that is ignored in the broad-beam (BB) algorithm.\cite{Petti 1992, Hong 1996}
The dose distribution will be formed with superposition of the pencil beams using kernel-convolution techniques with variations in algorithmic implementation, for example, in choice of the coordinate system, order of the multiple integrals, and numerical approximations, which greatly influence the accuracy, speed, complexity, and generality of the code.\cite{Mackie 1985, Ahnesjoe 1999}

Though the PB algorithm may be sufficiently accurate and fast for dose calculation in treatment planning in the present form, demand for faster calculation methods may always remain, for example, for optimization in the intensity-modulated radiotherapy with scanned charged particle beams,\cite{Soukup 2005, Lomax 2001} and for adaptive radiotherapy under image guidance,\cite{Yan 1997, Mackie 2003} which will ultimately accommodate on-site re-planning for an immobilized patient quickly between imaging and treatment.
Pursuing faster computational algorithms might be critical for the innovation to happen.

This paper presents one of such approaches, where we briefly review the BB and PB algorithms, describe the new algorithm, demonstrate the effectiveness in carbon-ion radiotherapy, evaluate the accuracy with a modeled proton pencil beam, and discuss the applicability for scanned beams.

\section{Materials and Methods}

\subsection{The broad-beam algorithm}
In the BB algorithm, dose $D$ at point $\vec{r}$ is resolved into the BB dose and the penumbra effect.\cite{Petti 1992, Hong 1996}
The BB dose $D_\mathrm{BB}$, or equivalently dose per fluence of the incident beam, is given either theoretically or experimentally as a function of water-equivalent depth $w$ that is calculated with the ray-tracing integral of effective density $\rho$ from the beam source $\vec{r}_0$ in radial direction $\vec{v} = (\vec{r}-\vec{r}_0)/ |\vec{r}-\vec{r}_0|$. 
 The penumbra effect gradates the field edge with the error function $\mathrm{erf}(x) = (2/\sqrt{\pi}) \int_0^x \mathrm{e}^{-u^2} \mathrm{d}u$ of the signed closest distance to the geometrical field edge, $t$ ($t > 0$ for $\vec{r}$ in the field, $t < 0$ otherwise),
\begin{eqnarray}
D(\vec{r}) &=& D_\mathrm{BB}\left(w(\vec{r})\right) \frac{1}{2} \left[1+\mathrm{erf}\!\left(\frac{t\left(\vec{r}\right)}{\sqrt{2}\,\sigma_\mathrm{t}(\vec{r})}\right)\right]
\label{eq_bba}
\\
w(\vec{r}) &=& \int_0^{\left|\vec{r}-\vec{r}_0\right| } \rho\left(\vec{r}_0+ s \, \vec{v}(\vec{r}) \right) \mathrm{d}s
\label{eq_wel}
\end{eqnarray}
where the projected transverse spread $\sigma_\mathrm{t}(\vec{r})$ is given either experimentally or theoretically.

The formulation of the penumbra effect simulates the dose-collecting process at point $\vec{r}$ from uniformly and continuously distributed invariant Gaussian sub-beams in the field.
The assumed uniformity and invariance of the sub-beams restrict the validity of the model to a narrow penumbra region where the local-homogeneity approximation may be valid.

\subsection{The pencil-beam algorithm}

In convolution algorithms, a dose distribution is generally calculated  by kernel integral,
\begin{equation}
D(\vec{r}) = \int T(\vec{p}) \, h(\vec{p},\vec{r}) \, \mathrm{d}^3\vec{p},
\label{eq_pba}
\end{equation}
where $T(\vec{p})$ is the total energy released per mass (terma) from the radiation at point $\vec{p}$ and kernel function $h(\vec{p},\vec{r})$ is the terma fraction transfered to point $\vec{r}$.\cite{Ahnesjoe 1999} 
In the PB algorithm, the terma equals the BB dose in the beam field or zero otherwise and is transversely spread by a planar Gaussian kernel,
\begin{eqnarray}
T(\vec{p}) &=& \begin{cases} D_\mathrm{BB}\left(w(\vec{p})\right) & \left(\vec{p} \in \text{field}\right) \\*
0 & \left(\vec{p} \notin \text{field}\right) \end{cases}
\label{eq_terma}
\\
h(\vec{p},\vec{r}) &=& \frac{1}{2 \, \pi \, \sigma_\mathrm{t}^2(\vec{p})} \, \mathrm{e}^{-\frac{|\vec{r}-\vec{p}|^2} {2 \, \sigma_\mathrm{t}^2(\vec{p}) }} \delta\left((\vec{r}-\vec{p}) \cdot \vec{v}(\vec{p}) \right), 
\label{eq_pbk}
\end{eqnarray}
where the Dirac $\delta$ function restricts the spreading in the plane perpendicular to the PB direction $\vec{v}$, leading the convolution to an areal integral of a pencil kernel in the field.\cite{Petti 1992, Hong 1996}
In the numerical integration, several tens or more terma-emitting points are usually arranged around each of the dose-collecting points on the transverse plane with radial distance limitation $|\vec{r}-\vec{p}| < \alpha\, \sqrt{2}\, \sigma_\mathrm{t}$, where the Gaussian tail-cutoff parameter $\alpha$ is normally set to 3 and a normalization factor is multiplied to the kernel to compensate the ignored tail contributions.\cite{Hong 1996}

The PB algorithm accommodates the density heterogeneity by involving the ray-tracing integral Eq.~(\ref{eq_wel}) to derive the terma and the kernel within the convolution integral Eq.~(\ref{eq_pba}).
The multiple integration will, however, increase the computational amount severely.

\subsection{The grid-dose-spreading algorithm}\label{II.C}

In treatment planning, the dose grids must be fine enough to show dose variation in the patient with grid spacing as small as $\sigma_\mathrm{t}$ or less and should be also able to represent distributions of any quantities.
In the grid-dose-spreading (GDS) algorithm, the termas and the spreads in Eqs.~(\ref{eq_terma}) and (\ref{eq_pbk}) are calculated at all the dose grids and stored in three-dimensional arrays,
\begin{eqnarray}
T_i &=& T(\vec{r}_i) \label{eq_ti}
\\
{\sigma_\mathrm{t}}_i &=& \sigma_\mathrm{t}(\vec{r}_i),
\end{eqnarray}
for grid $i$ located at $\vec{r}_i = (x_i, y_i, z_i)$ in the grid-based coordinate system.
The number of the ray-tracing integrals is minimized by extracting out of the convolution integral.

The gridded distributions, however, are not directly applicable to the convolution due to the coplanar constraint between terma emission and dose collection in the PB model because the PB axis is generally angled to the $x$, $y$, and $z$ grid axes with direction cosine vector $\vec{v}=(v_\mathrm{x},v_\mathrm{y},v_\mathrm{z})$.
In order to resolve this difficulty, the disk-shaped kernel in Eq.~(\ref{eq_pbk}) is deformed to the best approximate ellipsoidal kernel of the product of three Gaussian functions, 
\begin{eqnarray}
h(\vec{p},\vec{r}) &\to& \frac{\mathrm{e}^{-\frac{q_\mathrm{x}^2} {2 \, \sigma_\mathrm{x}^2}}}{\sqrt{2 \, \pi} \, \sigma_\mathrm{x}} \, 
\frac{\mathrm{e}^{-\frac{q_\mathrm{y}^2} {2 \, \sigma_\mathrm{y}^2}}}{\sqrt{2 \, \pi} \, \sigma_\mathrm{y}} \,
\frac{\mathrm{e}^{-\frac{q_\mathrm{z}^2} {2 \, \sigma_\mathrm{z}^2}}}{\sqrt{2 \, \pi} \, \sigma_\mathrm{z}} ,
\end{eqnarray}
where $\vec{q} = \vec{r}-\vec{p} = (q_\mathrm{x},q_\mathrm{y},q_\mathrm{z})$ is the displacement vector from the terma-emitting to dose-collecting points and $\sigma_\mathrm{x}$, $\sigma_\mathrm{y}$, and $\sigma_\mathrm{z}$ are the grid-axial projections of the spread, 
\begin{eqnarray}
\sigma_k^2(\vec{p}) &=& \int q_k^2 \, h(\vec{p}, \vec{p}+\vec{q}) \, \mathrm{d}^3\vec{q} \nonumber\\*
&=& \left(1-v_k^2(\vec{p})\right) \, \sigma_\mathrm{t}^2(\vec{p}) \quad (k = \text{x, y, z}),
\end{eqnarray}
derived with the planar point kernel in Eq.~(\ref{eq_pbk}).
Applying the gridded distributions, the convolution in the PB algorithm in Eq.~(\ref{eq_pba}) is rewritten to
\begin{equation}
D_j = \!\!\!\!\!\! \sum_{ i \in \left( \left| {q_k}_{ij} \right| \leq \alpha\, \sigma_k + \frac{\delta_k}{2} \right)_{\forall k}} \!\!\!\!\!\! T_i \, \prod_k \frac{{h_k}_i ({q_k}_{ij})}{{\epsilon_k}_i}
\end{equation}
where $D_j$ is the dose collected at grid $j$, $\vec{q}_{ij}$ is the displacement between grids $i$ and $j$, grid-axial spreading function ${h_k}_i({q_k}_{ij})$ is the dose fraction transfered into width $\delta_k$ at grid $j$, and dose-collection acceptance ${\epsilon_k}_i$ compensates the ignored Gaussian tails by cutoff parameter $\alpha$ for the summation.
The grid-axial spreading functions and the acceptances are analytically given by
\begin{eqnarray}
{h_k}_i(q) &=& \frac{1}{\sqrt{2\,\pi}\, {\sigma_k}_i } \int_{q-\frac{\delta_k}{2}}^{q+\frac{\delta_k}{2}} \mathrm{e}^{-\frac{u^2}{2\, {\sigma_k}_i^2}}\, \mathrm{d}u 
\nonumber\\*
&=&\frac{1}{2} \left[ \mathrm{erf}\!\left( \frac{|q|+\frac{\delta_k}{2}} {\sqrt{2}\,{\sigma_k}_i} \right) - \mathrm{erf}\!\left(\frac{|q|-\frac{\delta_k}{2}}{\sqrt{2}\,{\sigma_k}_i} \right) \right]
\\
{\epsilon_k}_i &=& \frac{1}{\sqrt{2\,\pi}\, {\sigma_k}_i } \int_{-\alpha\, {\sigma_k}_i -\frac{\delta_k}{2}}^{\alpha\, {\sigma_k}_i +\frac{\delta_k}{2}} \mathrm{e}^{-\frac{u^2}{2\, {\sigma_k}_i^2}}\, \mathrm{d}u
\nonumber\\*
 &=& \mathrm{erf}\!\left(\frac{\alpha\, {\sigma_k}_i+\frac{\delta_k}{2}}{\sqrt{2}\, {\sigma_k}_i} \right),
\end{eqnarray}
which can be quickly computed with the standard math library.

Since the effective kernel volume is conserved in the deformation, the computational amount is proportional to the cutoff cross section of the pencil beam or roughly to $\alpha^2$.
The computational efficiency is maximized by adopting the convolution scheme so-called ``the interaction point of view''.\cite{Mackie 1985}
In this case, the acceptance-corrected terma $T_i/({\epsilon_\mathrm{x}}_i\,{\epsilon_\mathrm{y}}_i\,{\epsilon_\mathrm{z}}_i)$ is calculated at each emitter grid $i$ and then is distributed with the fractions $\prod_k {h_k}_i({q_k}_{ij})$'s to the nearby grids $j$'s. 
The gridded dose distribution $D_j$ will be formed when all the terma emissions have been processed as above.

It is noted that the GDS and PB algorithms may share all the physical and computational models and their inaccuracies except for the grid quantization and the kernel deformation.
In cases where these additional inaccuracies are substantially smaller than the other ones, the GDS and PB algorithms will be practically equivalent.

\subsection{Implementation to treatment planning system for carbon-ion radiotherapy}\label{II.D}

Carbon-ion radiotherapy has been practiced at National Institute of Radiological Sciences since 1994 with accelerator complex HIMAC,\cite{Kanai 1999} and original treatment planning system HIPLAN,\cite{Endo 1996, Kanematsu 2002} where the BB algorithm has been consistently used to avoid disturbance to the ongoing clinical studies even though Petti and Kohno {\it et al.} explicitly showed that the BB algorithm involves principled inaccuracy due to lack of beam blurring in the field.\cite{Petti 1992, Kohno 2004}

The GDS algorithm as described in Sec.~\ref{II.C} was implemented to HIPLAN using the existing framework of the BB code.
The terma distribution $T(\vec{r})$ is calculated by applying $\sigma_\mathrm{t} \to 0$ in Eq.~(\ref{eq_bba}), where the depth--dose curve $D_\mathrm{BB}(w)$ is from the beam data library of HIPLAN,\cite{Kanematsu 2007} for the range-modulated carbon-ion beams including relative biological effectiveness (RBE) correction.\cite{Kanai 1999}
Invariant $\sigma_\mathrm{t} = 4$ mm and $\alpha = 3$ are used to preserve the penumbra behavior of the BB algorithm for the $400$-MeV/nucleon beams with the multileaf collimator. 

Generally, two algorithms can be impartially compared only under exactly the same condition  except for the essential algorithmic differences in implementation.
In this regard, we can accurately compare the GDS and BB algorithms by applying them to the identical plan on the single HIPLAN system.

\subsection{Analytic proton pencil-beam model}\label{II.E}

Since the GDS algorithm is a variant of the PB algorithms with additional approximations, it is necessary and sufficient for the accuracy evaluation to examine how the GDS calculation reproduces a modeled pencil beam under realistic conditions in grid spacing, transverse spread, and incident angle.
For the PB model, a proton beam in water, or the proton pencil kernel itself, may be the most appropriate because the spread may be the largest among the ion species and the errors will be the clearest for simplicity. 

Bortfeld established an analytic model for the proton Bragg curve in water, $D_\mathrm{BB}(w)$ in our notation, as
\begin{eqnarray}
D_\mathrm{BB}(w) =&& \frac{\mathrm{e}^{-\case{(R_0-w)^2}{4\,\sigma_\mathrm{R}^2}} \sigma_\mathrm{R}^{0.565}}{1+0.012\,R_0} 
\left[ \frac{11.26}{\sigma_\mathrm{R}}\,\mathcal{D}_{-0.565}\!\!\left(-\case{R_0-w}{\sigma_\mathrm{R}}\right) \right. \nonumber \\
&& \left. +0.157\,\mathcal{D}_{-1.565}\!\!\left(-\case{R_0-w}{\sigma_\mathrm{R}}\right) \right],
\end{eqnarray}
in MeV\,g$^{-1}\,$cm$^2$ per incident proton,\cite{Bortfeld 1997} where $\mathcal{D}_n(u)$ is the parabolic cylinder function,\cite{Miller 1972} and $R_0$ and $\sigma_\mathrm{R}$ are the beam range and the range straggling in g/cm$^2$ related to incident energy $E_0$ in MeV with formulas $R_0=0.0022\,{E_0}^{1.77}$ and $\sigma_\mathrm{R}=0.012\,{R_0}^{0.935}$, respectively.\cite{Bortfeld 1997} 

Hong {\it et al.} tabulated the projected transverse scatter of protons in water,\cite{Hong 1996} which is approximated by function $y_0(w)=0.023\,w\,(0.83\,w/R_0+0.17)$ with accuracy better than a fraction of a millimeter.
For a pencil beam generated at $\vec{r}_0$ with projected angular spread $\theta_0$, the divergent term $\theta_0\, s$ proportional to distance $s$ is quadratically added to the scatter,
\begin{equation}
 {\sigma_\mathrm{t}}^2(s) = 0.023^2\,w^2(s)\left(0.83\,\frac{w(s)}{R_0}+0.17\right)^2+{\theta_0}^2\,s^2,
\end{equation}
where $w(s) = \int_0^s \rho\left(\vec{r}_0+ s'\,\vec{v}\right) \mathrm{d}s'$ is the water-equivalent depth from $\vec{r}_0$.
The dose distribution for a proton pencil beam in water, $D(\vec{r})$, is then given by
\begin{equation}
D(\vec{r}) = D_\mathrm{BB}(w(\vec{r})) \, \frac{1}{2 \pi \sigma_\mathrm{t}^2(s(\vec{r}))} \, \mathrm{e}^{-\case{|\vec{r}-\vec{r}_0|^2-s^2(\vec{r})}{2 \sigma_\mathrm{t}^2(s(\vec{r}))}},
\end{equation}
where $s(\vec{r}) = \vec{v} \cdot (\vec{r}-\vec{r}_0)$ is the projected distance from $\vec{r}_0$ to $\vec{r}$ along $\vec{v}$.

In applying the GDS algorithm to this system,  the gridded spread is derived by ${\sigma_\mathrm{t}}_i = \sigma_\mathrm{t}(s(\vec{r}_i))$, while the gridded terma $T_i$ is calculated according to the definition as total energy released per mass in the grid-$i$ voxel, 
\begin{equation}
T_i = \frac{1}{V_0} \int_{{s_\mathrm{in}}_i}^{{s_\mathrm{out}}_i}\!\!\! D_\mathrm{BB}\left(w(s)\right)\,\mathrm{d}s,
\end{equation}
where $V_0$ is the volume of the voxel, and ${s_\mathrm{in}}_i$ and ${s_\mathrm{out}}_i$ are the distances on the beam axis to enter and to exit from the voxel, respectively, since Eq.~(\ref{eq_terma}) is not applicable to the infinitesimal field.
The subsequent formulation in Sec.~\ref{II.C} is then applicable to this system.

\section{Results}

\subsection{Performance in carbon-ion radiotherapy}\label{III.A}

\begin{figure}
\includegraphics[width=6.5cm]{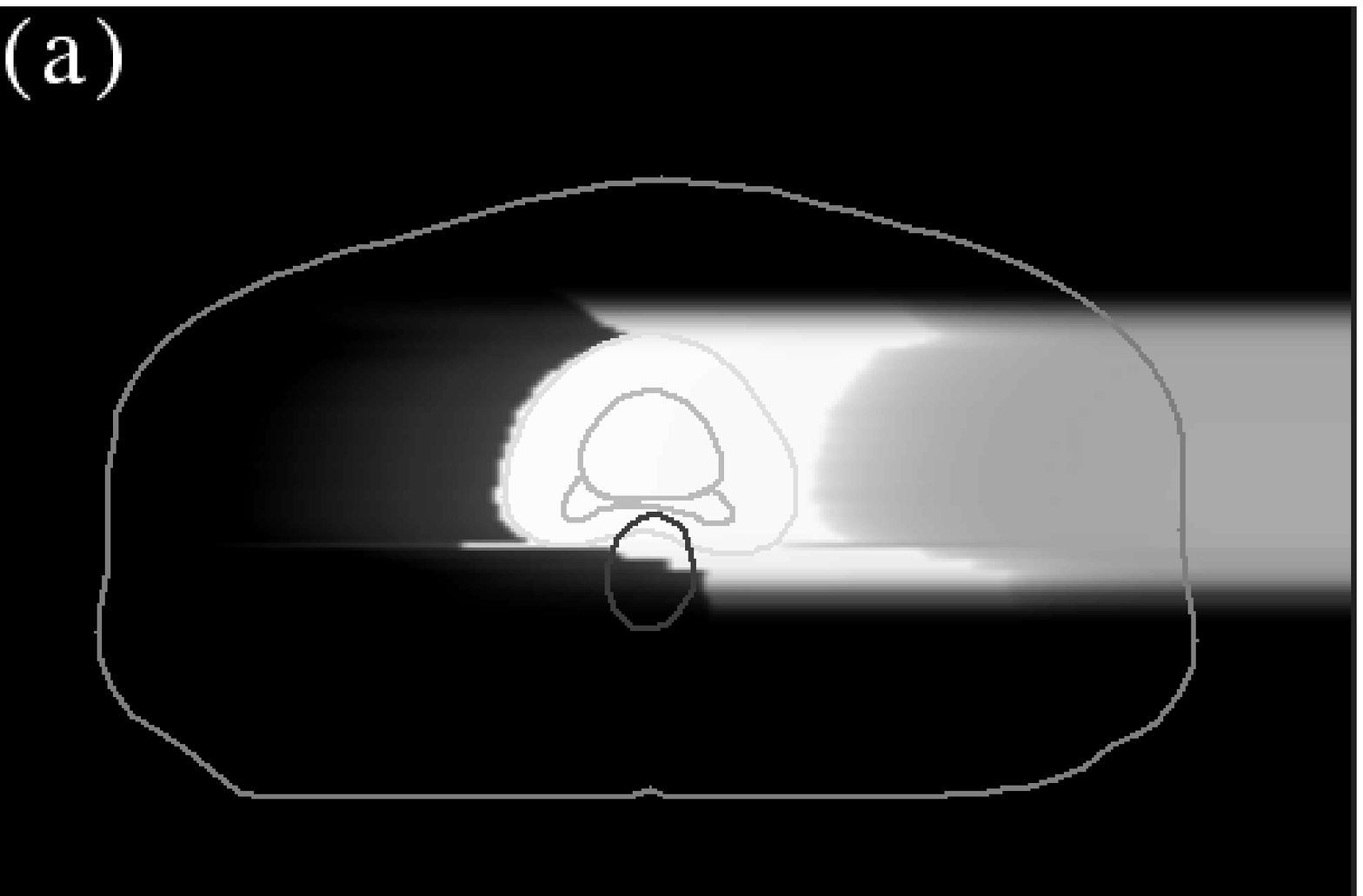}\\
\includegraphics[width=6.5cm]{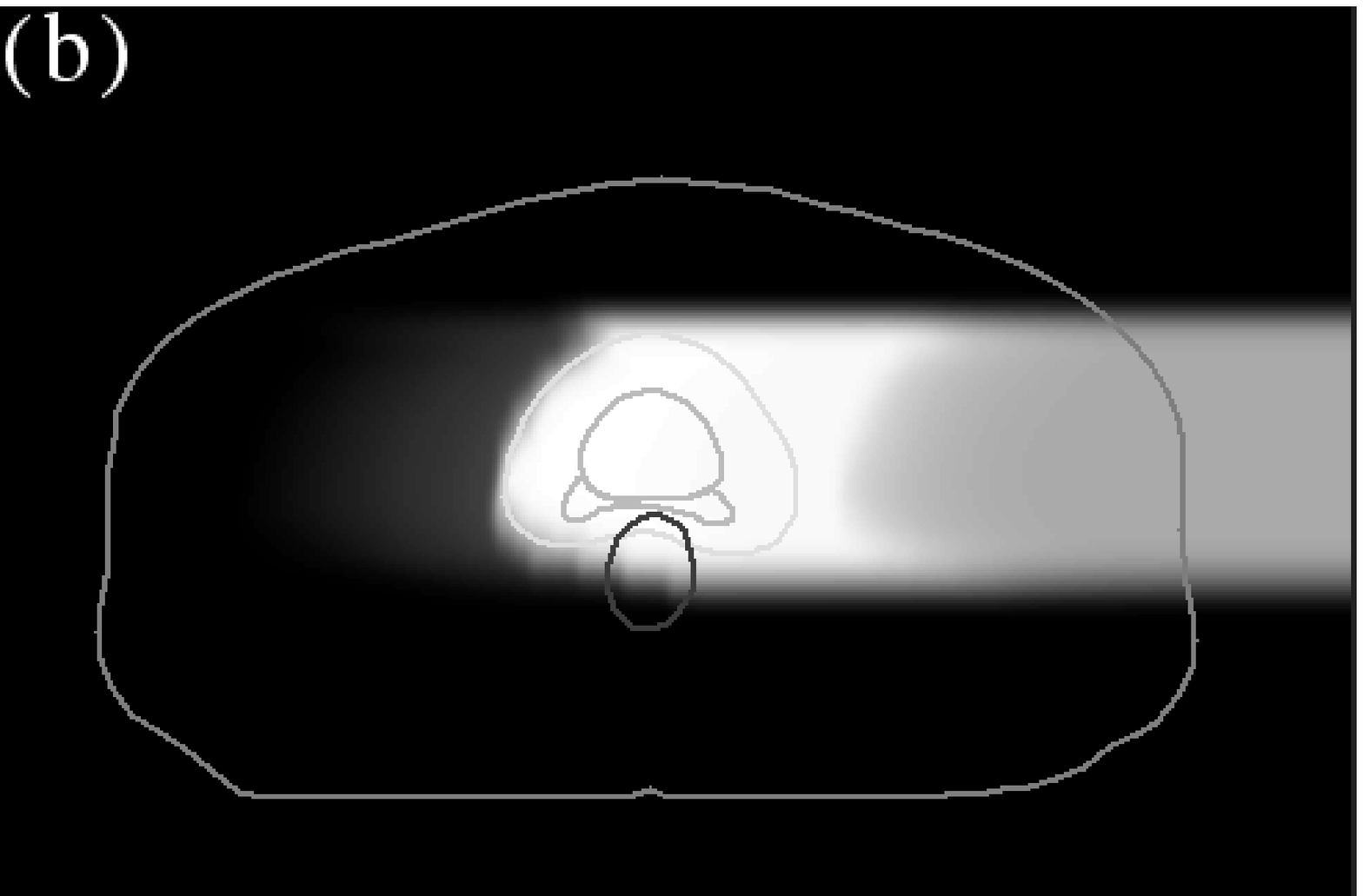}
\caption{Clinical dose distributions in a transaxial plane (43 cm $\times$ 28 cm) from a carbon-ion beam for prostate treatment calculated with the (a) BB and (b) GDS algorithms, where a prostate and seminal vesicles (gray lines) are included in the target (light gray) that partly overlaps with a rectum (dark gray) in a patient (gray), a horizontal beam is incident from the patient's left (the figure's right), and the doses are in linear gray scale (black for zero to white for the maximum).}
\label{fig_1}
\end{figure}

Figure \ref{fig_1} shows the clinical, or RBE-weighted, dose distributions of the GDS and BB calculations on HIPLAN for a clinical case of carbon-ion radiotherapy for prostate.
The clinical target volume (CTV) consisting of the prostate and the seminal vesicles, the rectum as an organ at risk, and the planning target volume (PTV) with 5-mm margin to the rectum side and 10-mm margin elsewhere added to the CTV, were manually segmented.\cite{Tsuji 2005}
The effective density distribution for heterogeneity correction was derived from the planning CT image,\cite{Kanematsu 2003} with grid spacings of 1.758 mm along the right--left and anterior--posterior axes and 2.500 mm along the inferior--superior axis, which are shared by the dose distributions.

A horizontal beam was conformed to the PTV with minimum 6 mm field margin using a multileaf collimator and was customized with a range compensator, a sculptured plastic object attached to the port, to absorb extra penetration of the carbon ions beyond the PTV with 3 mm depth margin.
The compensator was designed in the $3 \times 3$ mm$^2$-sized pixel-array format with steepness limited by the maximum depth step of 15 mm considering the tapered structure of the milling tool.
In this example, the rectum side of the PTV is almost parallel to the horizontal beam and the range compensation results in steep variation in beam range or so-called range discontinuity.

The BB calculation in Fig.~\ref{fig_1}(a) exhibits unphysically too sharp dose gradient at the range discontinuity around the rectum at risk, which could be influential on the clinical plan review.
The ripples and the spikes of the range surface came from incomplete range compensation within the $3 \times 3$-mm$^2$ pixels.
These artifacts have been naturally smeared out in the GDS calculation in Fig.~\ref{fig_1}(b).

While the BB algorithm was designed to reproduce the error function of $\sigma_\mathrm{t} = 4$ mm in the penumbra region, there was submillimeter-level disagreement between the BB and GDS calculations in field edge defined by 50\%-dose position, which is consistent with the grid-quantization error of the GDS algorithm.
The BB and GDS calculations for the prostate treatment case took 48 s and 66 s with SGI$^{\text{\textregistered}}$ Octane workstation, respectively. 
Namely, the GDS calculation was 1.4 times slower than the BB calculation in this case.

\subsection{Accuracy for angled proton pencil beam}\label{III.B}

\begin{figure}
\includegraphics[width=8.5cm]{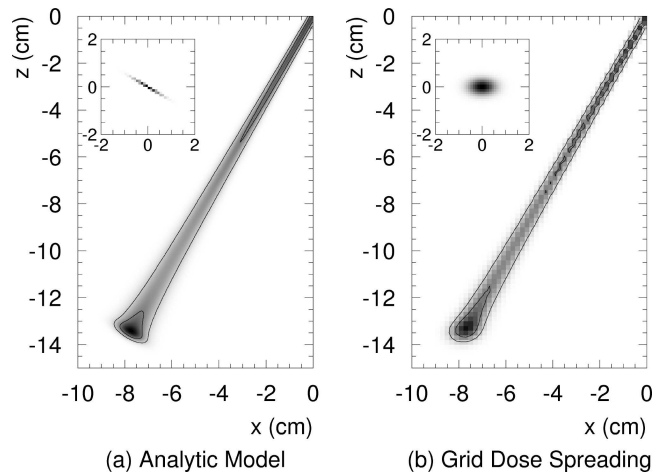}
\caption{Dose distributions from a proton pencil beam in water projected onto the $x$--$z$ plane, (a) the analytic beam model and (b) the corresponding GDS calculation. The $z$ axis is the vertical height from the water level and the $x$ axis is the relative horizontal position. The 20\% and 50\% isodose lines relative to the analytic maximum of 29.2 MeV\,g$^{-1}$\,cm per incident proton are drawn with the gray scale images. The embedded images show the point-spread functions.}
\label{fig_2}
\end{figure}

Figure \ref{fig_2} shows dose distributions in water projected onto the $x$--$z$ plane, where a point-like 150-MeV proton beam with projected angular spread $\theta_0 = 10$ mrad is generated at 10 cm above water level with zenith angle $30^\circ$, namely with $\vec{r}_0 = (10/\sqrt{3}, 0, 10)$ cm and $\vec{v} = (-1/2, 0, -\sqrt{3}/2)$ in the grid-based coordinate system with origin defined at the beam entrance point into water.
The grid spacings in the GDS calculation are all 2 mm along the three axes.

At relatively shallow depth, the GDS and the analytic model calculations are consistent within the grid resolution especially in the 20\% isodose line while the 50\% isodose line suffers from small dose errors under the low-dose-gradient condition.
At the Bragg peak, there is substantial disagreement in the 20\% isodose line within about 3 mm.

The embedded images in Fig.~\ref{fig_2} show the point-spreading functions with projected transverse spread of $\sigma_\mathrm{t} = 4.5$ mm at the Bragg peak, where the spreading in the analytic model is confined in the transverse plane and that in the GDS algorithm forms an ellipsoidal volume. 
In other words, the planar spreading is approximately resolved into the three uncorrelated orthogonal spreading by $\sigma_\mathrm{x} = 3.9$ mm, $\sigma_\mathrm{y} = 4.5$ mm, and $\sigma_\mathrm{z} = 2.3$ mm, which has deformed the point-spreading function, or the point kernel, and consequently the dose distribution.

\section{Discussion}

\subsection{Speed}

The PB algorithms superpose the dose distributions from the pencil beams in the field, where time-consuming ray-tracing integrals will severely slow down the calculation if performed within the convolution.
The algorithmic implementation from the interaction point of view,\cite{Mackie 1985} where multiple dose depositions are associated with each terma emission, may minimize the ray-tracing integrals.
However, the dose transfer to the nearby dose grids involves relatively complicated coordinate transformations within the convolution, which may be still substantially time consuming.\cite{Kanematsu 1998} 

In the GDS algorithm, the superposition is done in the terma space rather than in the dose space, the ray-tracing integrals are completely decoupled from the convolution, and the geometrical emitter--collector associations are maximally simplified to the grid-to-grid basis, which in all make the GDS algorithm very efficient.
In fact, Hong {\it et al.} studied the BB and PB algorithms for proton radiotherapy and found that the PB calculation was slower than the BB calculation by factor of 73 with a test case.\cite{Hong 1996}
The corresponding slowing factor for the GDS algorithm was only 1.4 in our prostate treatment case, indicating significant speed improvement over the conventional PB algorithms.

\subsection{Accuracy}

In the GDS algorithm, the terma distribution is calculated as an intermediate quantity with grid quantization errors, which will propagate to the dose distribution. 
However, the quantization error will be usually negligible with sufficiently fine grid spacing.
For example, in the prostate treatment case with grid spacing of 1.758 mm, the rms error from the quantization will be as small as $1.758/\sqrt{12} \approx 0.5$ mm, which is by far better than the realistic accuracy in target delineation.

\begin{figure}
\includegraphics[width=8.5cm]{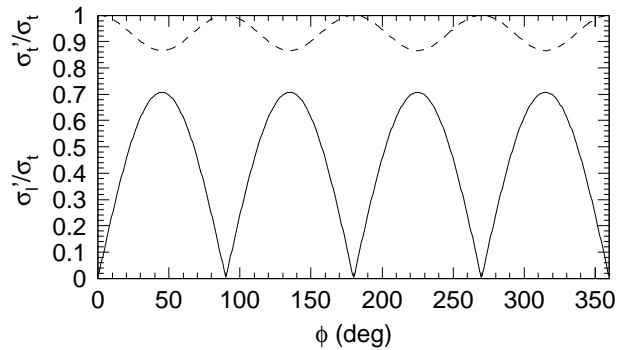}
\caption{The longitudinal ($\sigma_\mathrm{l}'$; solid line) and transverse ($\sigma_\mathrm{t}'$; dashed line) spreads of a coplanar beam reprojected from the deformed kernel in the GDS algorithm relative to the original transverse spread ($\sigma_\mathrm{t}$) as a function of gantry angle ($\phi$).}
\label{fig_3}
\end{figure}

The kernel deformation observed in the proton pencil beam case will greatly depend on beam direction with respect to the grid axes. 
When the beam is angled to all the three grid axes, the transverse planar spreading is approximated by volumetric spreading that includes an artifactual longitudinal component.
The longitudinal spread $\sigma_\mathrm{l}'$ and the transverse spread $\sigma_\mathrm{t}'$ of the deformed kernel are derived from reprojection and conservation of the spread squared,
\begin{eqnarray}
\sigma_\mathrm{l}' &=& \sqrt{\sum_k \sigma_k^2 \, v_k^2} = \sqrt{1-v_\mathrm{x}^4-v_\mathrm{y}^4-v_\mathrm{z}^4}\,\sigma_\mathrm{t}
\\
\sigma_\mathrm{t}' &=& \sqrt{\sigma_\mathrm{t}^2 - \frac{\sigma_\mathrm{l}'^2}{2}} = \frac{1}{\sqrt{2}}\,\sqrt{1+v_\mathrm{x}^4+v_\mathrm{y}^4+v_\mathrm{z}^4}\,\sigma_\mathrm{t},
\end{eqnarray}
which are a measure of inaccuracy in distal fall off and a measure of accuracy in lateral penumbra, respectively, and will be both $\surd(2/3) \, \sigma_\mathrm{t}$ in the worst case with direction $\vec{v} = (\pm 1, \pm 1,\pm 1) / \sqrt{3}$.
Figure \ref{fig_3} shows the reprojected spreads as a function of gantry angle $\phi$ in coplanar beam arrangement with $v_\mathrm{y} \approx 0$, 
%with $v = (\sin \phi, 0, \cos \phi)$, or specifically $\sigma_\mathrm{l}' = \surd(2\,\sin^2\phi\,\cos^2\phi)\,\sigma_\mathrm{t}$ and $\sigma_\mathrm{t}' = \surd(1-\sin^2\phi\,\cos^2\phi)\,\sigma_\mathrm{t}$.
which explains the deformation of the proton pencil beam at $\phi = 30^\circ$ in Sec.~\ref{III.B}.

The spread of a pristine pencil beam limits the granularity of the dose distribution and may naturally approximate the necessary and sufficient spatial resolution for beam control and dose evaluation.
The finer structure below the resolution and the various spatial uncertainties are normally tolerated with appropriate margins.
In fact, the PTV should include substantial margins against patient setup error and internal organ motion of typically a few to several millimeters,\cite{Langen 2001} for example 5 to 10 mm for the prostate in Sec.~\ref{III.A}.
For clinical proton beams, the 20\%--80\% penumbra size may grow as large as 10 mm,\cite{Hong 1996} or $\sigma_\mathrm{t} \approx 10/1.68 \approx 6$ mm, against which, a field margin of $1.5\, \sigma_\mathrm{t} \sim 2\, \sigma_\mathrm{t} \approx 10$ mm around the PTV is usually added.
Then, even with the worst direction $\vec{v} = (\pm 1, \pm 1,\pm 1) / \sqrt{3}$, the artifact as large as $\sigma_\mathrm{l}' = \surd(2/3)\, \sigma_\mathrm{t} \approx 5$ mm will be mostly covered up by those margins.

Generally for broad beams, the systematic deformation of the kernels uniformly distributed in the field will be mostly compensated except for field edges in analogy to the kernel-tilting approximation for photon beams.\cite{Sharpe 1993}
The spread and hence the deformation will be even smaller with heavier-ion beams.
In addition, when a vertical or horizontal coplanar beam is used in conjunction with planning CT in treatment position, namely with $\phi$ = 0 or 90 in Fig.~\ref{fig_3}, the artifact will be completely absent, which has been almost always the case in carbon-ion radiotherapy with HIMAC and will be as well with its planned successors.\cite{Noda 2006}

There are also other sources of inaccuracy in the PB model,\cite{Kohno 2004} as well as ones in real clinical systems such as uncertainties in patient modeling, patient setup, organ motion, physiological variation, and beam control.
Since there are always requirements for practicalness in treatment planning, accuracy is important but not the absolute measure of dose calculation and should be assessed in a relative manner.

\subsection{Applicability}

The implementation of the GDS algorithm to the treatment planning system HIPLAN should have proved its simplicity, effectiveness, usability, and robustness necessary for dose distribution calculation in treatment planning practice and may also indicate possibilities for broader applications in the future.

For example, the framework for the single pencil beam in Sec.~\ref{II.E} can be extended to scanned beams, where the pencil beams are dynamically modulated in intensity, position, range, and spread.\cite{Schaffner 1999, Kramer 2000, Soukup 2005}
The instantaneous pencil beam, or spot beam $j$, is characterized by dose-per-fluence-per-incident-particle distribution ${D_\mathrm{BB}}_j(w)$, number of incident particles $N_j$, and transverse spread ${{\sigma_\mathrm{t}}_{j}}_i$ at grid $i$. 
The effective terma $T_i$ and the effective transverse spread ${\sigma_\mathrm{t}}_i$ at grid $i$ are calculated by dose-weighted superposition of the spot beams,
\begin{eqnarray}
T_i &=& \frac{1}{V_0} \sum_{j} \int_{{{s_j}_\mathrm{in}}_i}^{{{s_j}_\mathrm{out}}_i}\!\!\! N_j \, {D_\mathrm{BB}}_j\left(w(s_j)\right)\,\mathrm{d}s_j
\\
{\sigma_\mathrm{t}}_i^2 &=& \frac{1}{V_0 \, T_i} \sum_{j} \int_{{{s_j}_\mathrm{in}}_i}^{{{s_j}_\mathrm{out}}_i}\!\!\! {{\sigma_\mathrm{t}}_{j}}_i^2 \, N_j \, {D_\mathrm{BB}}_j\left(w(s_j)\right)\,\mathrm{d}s_j,
\end{eqnarray}
where parameter $s_j$ is the distance along spot beam $j$ and ${{s_j}_\mathrm{in}}_i$ and ${{s_j}_\mathrm{out}}_i$ are the distances to enter and to exit from the grid-$i$ voxel.
These gridded distributions will be calculated efficiently from the interaction point of view,\cite{Mackie 1985} for the GDS algorithm in Sec.~\ref{II.C}.
In other words, in the GDS algorithm, summation of the pencil beams per field is efficiently made in the terma and spread spaces and then a single volumetric convolution is also efficiently applied to form the dose distribution.

Application to intensity-modulated particle-beam therapy to achieve uniform target dose with multidirectional nonuniform fields,\cite{Soukup 2005, Lomax 2001} however, requires some caution because the kernel deformation of the GDS algorithm is orientation dependent and, thus, will not be compensated in the summed plan-dose distribution even in the middle of the treated volume.
Appropriateness of the beam modulation will have to be assessed, or otherwise compromised, to be less sensitive to not only the kernel deformation but also all the other uncertainties involved in the real clinical system.\cite{Lomax 2001} 

\section{Conclusions}

A new variant of the PB algorithm, the GDS algorithm, is proposed for heavy charged particle radiotherapy with approximations of the gridded intermediate distributions and a modified convolution kernel for grid-to-grid energy transfer.
The resultant high-speed nature and easiness of implementation are distinctive features of the GDS algorithm.

When the beam incidence is angled to all the dose-grid axes, the approximation will cause deformation in dose distribution within the size of the pristine spread.
Such inaccuracy will have to be assessed relatively against the clinical tolerances and the other sources of errors in practical situations.

\end{document}